\documentclass[twocolumn,superscriptaddress,aps,prl,floatfix]{revtex4-2}
\usepackage{graphicx}
\usepackage{dcolumn}
\usepackage{bm}
\usepackage[colorlinks=true,linkcolor=blue,filecolor=blue,urlcolor=blue,citecolor=blue]{hyperref}
\usepackage{gensymb}
\usepackage{footnote}
\usepackage[T1]{fontenc}
\begin{document}
\title{First observation of multi-phonon $\gamma$-vibrations in an odd-odd nuclear system}

\author{E.H. Wang}
\email{ehwang@sdu.edu.cn}
\affiliation{Shandong Provincial Key Laboratory of Nuclear Science, Nuclear Energy Technology and Comprehensive Utilization, Weihai Frontier Innovation Institute of Nuclear Technology, School of Nuclear Science, Energy and Power Engineering, Shandong University, Jinan 250061, China}
\affiliation{Weihai Research Institute of Industrial Technology of Shandong University, Weihai 264209, China}
\affiliation{Department of Physics and Astronomy, Vanderbilt University, Nashville, Tennessee 37235, USA}
\author{M. Abushawish}
\affiliation{Universit\'{e} Claude Bernard Lyon 1, CNRS/IN2P3, IP2I Lyon, UMR 5822, Villeurbanne, F-69100, France}
\author{J.H. Hamilton}
\affiliation{Department of Physics and Astronomy, Vanderbilt University, Nashville, Tennessee 37235, USA}
\author{A. Navin}
\email{navin.alahari@ganil.fr}
\affiliation{GANIL, CEA/DRF-CNRS/IN2P3, BP 55027, 14076 Caen cedex 5, France}
\author{S. Bhattacharyya}
\affiliation{Variable Energy Cyclotron Centre, 1/AF Bidhannagar, Kolkata 700064, India}
\affiliation{Homi Bhabha National Institute, Training School Complex, Anushaktinagar, Mumbai-400094, India}
\author{J. Dudouet}
\affiliation{Universit\'{e} Claude Bernard Lyon 1, CNRS/IN2P3, IP2I Lyon, UMR 5822, Villeurbanne, F-69100, France}
\author{G. H. Bhat}
\affiliation{Department of Physics, Government Degree College Shopian, Jammu and Kashmir 192303, India}
\author{J.A. Sheikh}
\affiliation{Department of Physics, Islamic University of Science and Technology, Awantipora 192122, India}
\affiliation{Department of Physics, University of Kashmir, Srinagar, 190006, India}
\author{S. Jehangir}
\affiliation{Department of Physics, Government Degree College Kulgam, Jammu and Kashmir 192231, India}
\author{S.Y. Wang}
\email{sywang@sdu.edu.cn}
\author{S. Sun}
\author{B. Qi}
\affiliation{Shandong Provincial Key Laboratory of Nuclear Science, Nuclear Energy Technology and Comprehensive Utilization, Weihai Frontier Innovation Institute of Nuclear Technology, School of Nuclear Science, Energy and Power Engineering, Shandong University, Jinan 250061, China}
\affiliation{Weihai Research Institute of Industrial Technology of Shandong University, Weihai 264209, China}
\author{M. Rejmund}
\affiliation{GANIL, CEA/DRF-CNRS/IN2P3, BP 55027, 14076 Caen cedex 5, France}
\author{A. Lemasson}
\affiliation{GANIL, CEA/DRF-CNRS/IN2P3, BP 55027, 14076 Caen cedex 5, France}
\author{Y. H. Kim} 
\altaffiliation{Present address: Center for Exotic Nuclear Studies, Institute for Basic Science, Daejeon 34126, Republic of Korea}
\affiliation{GANIL, CEA/DRF-CNRS/IN2P3, BP 55027, 14076 Caen cedex 5, France}
\author{E. Cl\'ement}
\affiliation{GANIL, CEA/DRF-CNRS/IN2P3, BP 55027, 14076 Caen cedex 5, France}
\author{F. Didierjean}
\affiliation{Universit\'{e} de Strasbourg, CNRS, IPHC UMR 7178, F-67000 Strasbourg, France}
\author{R.Y. Dong}
\affiliation{School of Physics and Astronomy, Sun Yat-sen University, Zhuhai 519082, China}
\author{G. Duch\^{e}ne}
\affiliation{Universit\'{e} de Strasbourg, CNRS, IPHC UMR 7178, F-67000 Strasbourg, France}
\author{B.~Jacquot}
\affiliation{GANIL, CEA/DRF-CNRS/IN2P3, BP 55027, 14076 Caen cedex 5, France}
\author{C.F. Jiao}
\affiliation{School of Physics and Astronomy, Sun Yat-sen University, Zhuhai 519082, China}
\author{Y.X. Luo}
\affiliation{Lawrence Berkeley National Laboratory, Berkeley, California 94720, USA}
\author{C. Michelagnoli}
\altaffiliation{Present address: Institut Laue-Langevin, F-38042 Grenoble Cedex, France}
\affiliation{GANIL, CEA/DRF-CNRS/IN2P3, BP 55027, 14076 Caen cedex 5, France}
\author{A.V. Ramayya}
\altaffiliation{Deceased}
\affiliation{Department of Physics and Astronomy, Vanderbilt University, Nashville, Tennessee 37235, USA}
\author{J.O. Rasmussen}
\affiliation{Lawrence Berkeley National Laboratory, Berkeley, California 94720, USA}
\author{C. Schmitt}
\altaffiliation{Present address: IPHC Strasbourg, Universit\'e de Strasbourg-CNRS/IN2P3, F-67037 Strasbourg Cedex 2, France}
\affiliation{GANIL, CEA/DRF-CNRS/IN2P3, BP 55027, 14076 Caen cedex 5, France}
\author{O. Stezowski}
\affiliation{Universit\'{e} Claude Bernard Lyon 1, CNRS/IN2P3, IP2I Lyon, UMR 5822, Villeurbanne, F-69100, France}
\author{W.Z. Xu}
\author{H. Zhang}
\affiliation{Shandong Provincial Key Laboratory of Nuclear Science, Nuclear Energy Technology and Comprehensive Utilization, Weihai Frontier Innovation Institute of Nuclear Technology, School of Nuclear Science, Energy and Power Engineering, Shandong University, Jinan 250061, China}
\affiliation{Weihai Research Institute of Industrial Technology of Shandong University, Weihai 264209, China}
\author{S.J. Zhu}
\affiliation{Department of Physics, Tsinghua University, Beijing 100084, China}

\date{\today}

\begin{abstract}
  The identification of the first multi-phonon $\gamma$-vibrational bands in an odd-odd neutron-rich nucleus of the nuclear chart is presented. These high spin structures of hard to access $^{104}_{41}$Nb$_{63}$, produced in fission, were studied by combining a spectrometer with isotopic resolution coupled to a $\gamma$-ray tracking array and independently high-fold $\gamma$ coincidence measurements.  Triaxial Projected Shell Model calculations for the high-spin states are in good agreement with the measured observables for the yrast, one-phonon and two-phonon $\gamma$ bands. The possibility of an oblate shape of an isomeric state and coexistence of triaxial and oblate configurations are investigated from the decay of the 141 keV isomer. The present work illustrates the robustness of vibration excitations in the presence of odd valence proton and neutron as well as the possibly coexisting shapes beyond the $N=60$ transitional region.
\end{abstract}
\maketitle
%


Among challenging problems in physics is the understanding of emergent collective excitation modes in interacting quantum
many-body systems in terms of the microscopic degrees of freedom. 
In atomic nuclei, the quadrupole field is the dominant collective excitation mode at low spin for a majority of atomic nuclei. 
The first $2^+$ excited band in most of the deformed and transitional even-even nuclei is due to one-phonon $\gamma$ vibrational mode in the collective model of Bohr and Mottelson \cite{Boh}. 
The collective model Hamiltonian is expressed in terms of $\beta$ and $\gamma$ deformation parameters with $\beta$ and $\gamma$ signifying the deviation from sphericity and axial symmetry (triaxial), respectively. 
The level energies of the one-phonon $\gamma$-vibrational band are indicative of $\gamma$-deformation and its softness \cite{Dav,Zam,Ani23}.

Microscopic investigations have provided new insights into the structure of the $\gamma$-bands in nuclei \cite{ys00,SJ21}. The angular-momentum projection from a triaxial state gives rise to a multiplet of band structures with well-defined values of $K$ (projection along the intrinsic symmetry axis) quantum number. From the triaxial vacuum (0 quasiparticle) state in even-even nuclei, the allowed values are $K=0,2,4,....$ \cite{ys00}. $K=0$ corresponds to the ground-state band, K=2 the first excited band with a band head of $2^+$ (the one-phonon $\gamma$-vibrational band or $\gamma$-band), K=4 with a band head of $4^+$ (two-phonon $\gamma$-vibrational band or the $\gamma\gamma$-band) and so on. From this microscopic perspective, $\gamma$-bands are built on each intrinsic state and $\gamma$-bands based on one- and three- quasiparticle configurations in odd-mass systems \cite{sj22}, and  two- quasiparticle states in even-even nuclei \cite{SJ18} have been identified. In odd-odd systems, multi-phonon collective excitation modes have not been identified yet. In most of the odd-odd nuclei, the energy spectrum is dominated by quasiparticle excitations and the observation of the $\gamma$-band as the first excited band would indicate the presence of the correlated ground-states.    

 \begin{figure*}
	\centering
	\includegraphics[width=\textwidth]{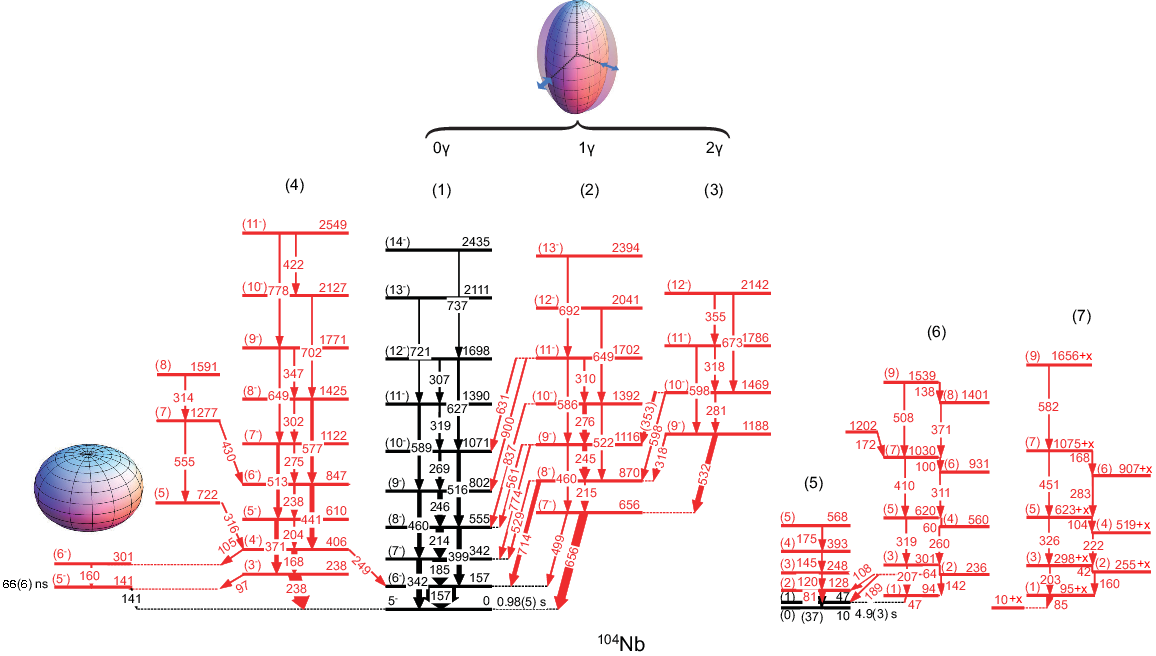}
	\caption{Level scheme of $^{104}$Nb. New transitions and levels are labeled in red. For the excitation energy of band 7, x is either 0 or 8.7 keV. Half-lives of the ground state and the 10 keV isomer are taken from NUBASE2020 \cite{Kon20}. }
	\label{level}  
\end{figure*} 

The study of multi-phonon excitations opens a window into the collective dynamics of many-body systems — whether they be atoms in a solid or nucleons in a nucleus. It highlights the nonlinear, correlated, and anharmonic nature of physical systems.  The multi-phonon $\gamma$-vibrational band was first observed in $^{168}$Er \cite{Bor91}, and then in other regions of A$\sim$100 \cite{Gue95}, $\sim$130 \cite{Li132}, and $\sim$230 \cite{Mar20}. These multi-phonon $\gamma$-vibrational bands have only been observed in even-even or even-odd nuclei. In case of odd-mass systems, the presence of the $\gamma$-band depends on the interplay between collective and single particle excitation modes, and fewer than ten cases have been reported \cite{Fry17}.

In neutron-rich nuclei around A$\sim$100, the first two-phonon $\gamma$-vibrational bands was observed in $^{106}_{42}$Mo \cite{Gue95}. Subsequently, similar band structures have been identified in $_{41}$Nb, $_{44}$Ru and $_{46}$Pd isotopes \cite{Gu10,Yeo11,Li13,Hua16,Fry17}. Thus, odd-odd Nb, Tc and Rh isotopes could be candidates for multi-phonon $\gamma$-vibrations and help understanding of the role of unpaired nucleons on collective  excitation modes. The fingerprints for multi-phonon $\gamma$-vibrational bands include the 2$\gamma$ band decaying only to 1$\gamma$ band and 1$\gamma$ band only to 0$\gamma$ band with enhanced E2 interband transitions \cite{Eld18}, similar moments of inertia \cite{Gue95}, energy differences between successive $\gamma$ phonon excitations \cite{Gue95},  and g factors (in the case of odd-mass nuclei) \cite{Wan09}.

$^{104}_{41}$Nb$_{63}$ lies in a region where rapid shape transitions including phenomena of a sudden onset of deformation at $N=60$ \cite{Ans17,Reg17}, shape coexistence \cite{Gar22,Gav232,Leo24,Kum23} and quantum phase transition \cite{Tog16, Gav22,Gav23} exist. Triaxial ground states have been reported in $N=60,62,64$ Nb isotopes \cite{Ha17} in comparison with the oblate ground state of $N=58$ Nb isotope \cite{Kum23}. Thus, this is an ideal region to understand how the valence proton and neutron drive the shape of the odd-odd nuclei especially for neutron-rich nuclei. The structure of $^{104}$Nb has been identified  by IAEA to be important for determining antineutrino energy spectra from nuclear reactors \cite{Dim15}.

In the case of odd-odd nuclei, a relatively more complex spectra and a low population of these bands, especially in neutron rich nuclei with small production cross sections, make it very challenging to identify multiphonon bands experimentally.
In this letter, a combination of two complimentary state-of-the-art spectroscopic
 tools was used to address the first observation of multi-phonon $\gamma$-vibrations in an odd-odd neutron-rich nucleus and a possible shape isomer in $^{104}$Nb.

The $^{104}$Nb nuclei were produced from spontaneous fission of $^{252}$Cf and fission using a $^{238}$U beam on a $^{9}$Be target.
The first measurement used a 62 $\mu$Ci $^{252}$Cf source in Gammasphere with 101 High Purity Germanium detectors at LBNL. The data were sorted into three- and four-fold $\gamma$-ray coincidence matrices and analyzed using the RADWARE package \cite{Rad}. More details of the experiment can be found in Ref.~\cite{Wan15}.
The measurements at GANIL were performed using a 6.2 MeV/u $^{238}$U beam on $^{9}$Be targets (1.6 $\mu$m and 5 $\mu$m) and two different values of the magnetic rigidity (B$\rho$) to optimize the yields of the fragments. The setup combined the VAMOS++ spectrometer \cite{Rej,Nav} for \textit{A, Z} identification of the fission fragments and the AGATA $\gamma$-ray tracking array \cite{Akk,Cle} to detect prompt $\gamma$ rays, and the EXOGAM array for delayed $\gamma$ rays \cite{Sim00}. The \textit{Z} was identified using a segmented ionization chamber at the focal plane of VAMOS++. The \textit{A} was obtained from the \textit{B$\rho$} and the measured velocity. An event-by-event Doppler correction was made using the velocity of the fragments and the corresponding $\gamma$-ray position in AGATA. The EXOGAM array consisting of seven segmented clover detectors was positioned behind the VAMOS++ ionization chamber to measure the delayed $\gamma$-rays in the range of 100 ns to 200 $\mu$s. Further experimental details are given in Ref.~\cite{Kim17,Lem24}. Results on the very neutron-rich odd-even isotopes $^{105,107,109}$Nb are reported elsewhere \cite{Abu24,Abu25}. 
 The above two complementary techniques combine the advantages of: a) unambiguously assignment of the $\gamma$ rays in 'singles' and $\gamma$-$\gamma$ coincidences in a $\gamma$ tracking spectrometer AGATA in conjunction with its full identification of (A,Z) using VAMOS++; b) high statistics triple and four-fold $\gamma$ coincidences to measure angular correlations, and identify the relatively weaker transitions. Complementary coincidence time windows also allow the identification of isomers from ns to $\mu$s.

The new level scheme of $^{104}$Nb with 7 band structures and other levels from the present work is shown in Fig.~\ref{level}. Bands (4)-(7) will be discussed elsewhere \cite{Wan25}. Band (1) was previously reported \cite{Wan08,Luo14} without linking transitions to the low-spin states observed in $\beta$-decay \cite{Rin07,Bla07}. A 130.5 keV isomeric transition with T$_{1/2}\gg 1\mu$s was reported in Ref.~\cite{Luo14} depopulating the band-head. Such an isomeric transition was not found in the present work and was attributed to the complementary fragment $^{146}$La \cite{Wan17}. In the delayed $\gamma$-ray spectrum, only the earlier reported 141 keV isomeric transition \cite{Joh70,Hop71,Wal70} was observed. 

\begin{figure}[htp!]
	\centering
	\includegraphics[width=\columnwidth]{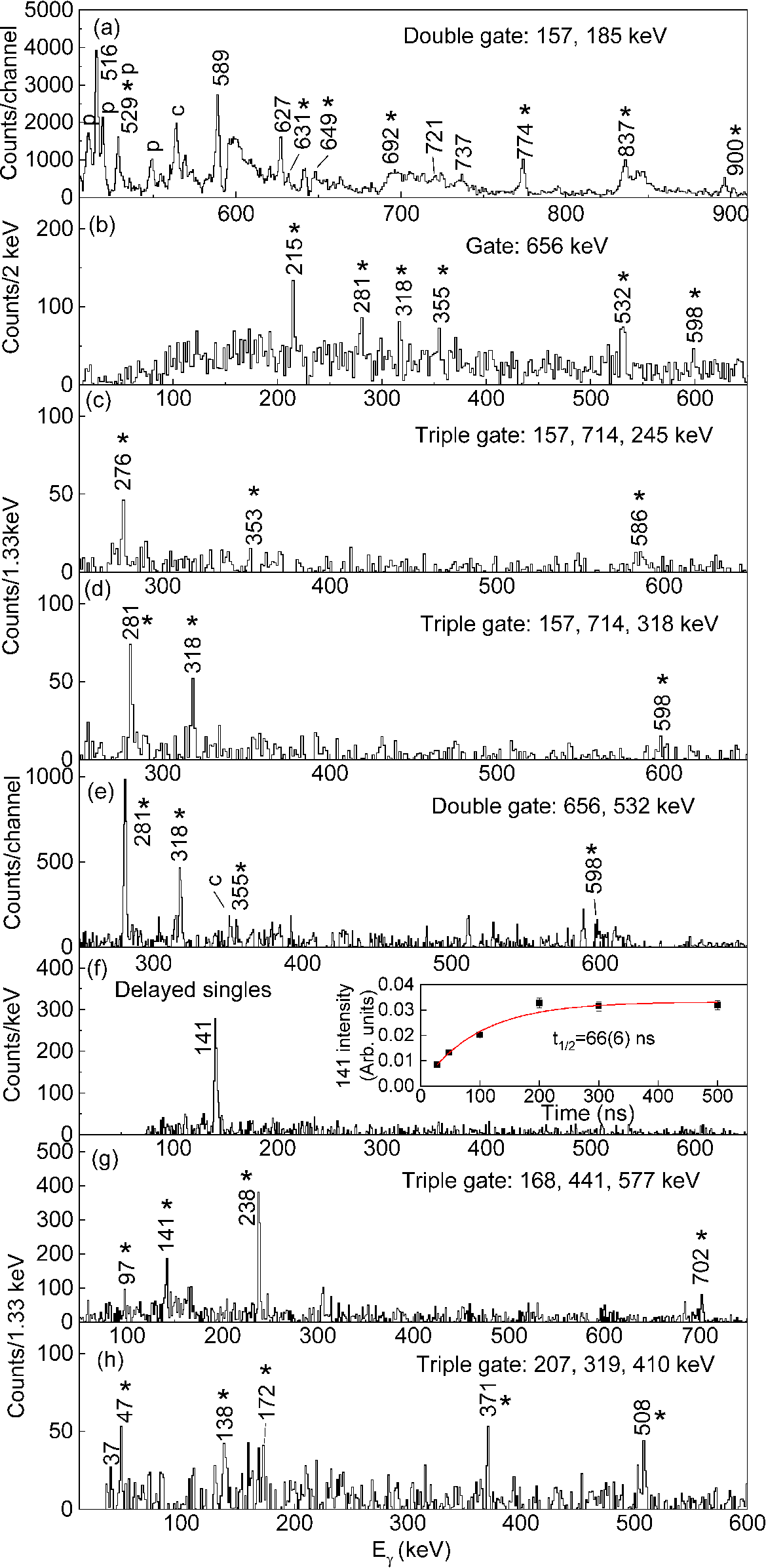}
	\caption{The $\gamma$-ray spectra from (a) gate on 157/185 keV;  (b) gate on 656 keV; (c) gate on 157/714/245 keV; (d) gate on 157/714/318 keV;   (e) gate on 656/532 keV; \color{black} (f) delayed singles, inset decay curve; (g) gate on 168/441/557 keV; (h) gate on 207/319/410 keV. The single gate and delayed singles are from $^{238}$U induced fission data, while the double and triple gates are from $^{252}$Cf data. New transitions are labeled with *. Fission partner La transitions are labeled with p. Contamination is labeled with c.}
	\label{spectra-all}   
\end{figure}

Coincidence spectra are shown in Fig.~\ref{spectra-all}. In part (a), the band (2) to band (1) linking transitions can be seen. Part (b) shows a gate on the strong prompt 656 keV transition linking the band-head of band (2) to band (1). Most of the strong transitions in bands (2) and (3), as well as the linking transitions from bands (3) to (2) can be observed. The coincidence spectra of Fig.~\ref{spectra-all}(c)/(d), show strong transitions in band (2)/band (3), respectively.  Those transitions are also confirmed by the 656 and 532 keV gate in Fig.~\ref{spectra-all}(e). In Fig.~\ref{spectra-all} (f), the delayed $\gamma$-ray singles of $^{104}$Nb are shown, where the 141 keV isomeric transition can be seen. Fig.~\ref{spectra-all}(g) shows evidence of the new transitions in band (4) and the transitions linking the 141 keV isomer. Fig.~\ref{spectra-all}(h) gives evidence of the transitions in the low spin band (6).

Previous $\beta$-decay \cite{Bla07,Rin07} and mass measurements \cite{Orf18,Huk23} showed the presence of a high-spin ground state with a T$_{1/2}$ of 0.98(5) s and a low-spin isomer with a half-life of 4.9(3) s at 10 keV in $^{104}$Nb. Band (1) in Fig.~\ref{level} is the yrast band, showing the strongest population in the two measurements, and thus the population of the high-spin ground state. A recent study of 
$^{104}$Nb $\beta$-decay \cite{Nan25} assigned 5$^-$ to the high-spin ground state, with a $\pi$5/2$^+$[422]$\otimes$$\nu$5/2$^-$[532] Nilsson configuration, based on an allowed transition from the high-spin ground state of $^{104}$Nb to the 2061 keV 4$^-$ level \cite{Mus21} of the daughter $^{104}$Mo. This assignment, in agreement with the 5/2$^-$ ground state in $^{103}$Zr \cite{Hua04,Urb09} and $^{105}$Mo \cite{Din06,Hua04}, and the 5/2$^+$ ground state in $^{103}$Nb \cite{Hua02,Luo05}, is adopted in the present work. 

The angular correlation of the intense 656–532 keV (band 2–3) cascade yields 
$A_2=0.09(3)$ and $A_4=0.02(5)$ and only the calculated $Q-Q$ values of $A_2=0.10$ and $A_4=0.01$ match the experiment, supporting spin-parity assignments of 7$^-$ and 9$^-$ for the band-heads of bands (2) and (3), respectively. The spins and parities of the other levels in $^{104}$Nb are tentatively assigned based on regular energy spacings and decay patterns. Thus, bands (1)–(4) are proposed to have high spins, corresponding to population of the high-spin 5$^-$ ground state, while bands (5)–(7) are proposed to have low spins, associated with population of the 10-keV low-spin isomer.

The new bands (2) and (3) are proposed to be the one- (1$\gamma$) and two-phonon (2$\gamma$) $\gamma$-vibrational bands coupled to band (1) with the reasons given
below. The proton orbitals near the Fermi surface are 5/2$^+$[422], 5/2$^-$[303] and 3/2$^-$[301] in $^{101,103,105,107,109}$Nb \cite{Hwa98,Luo05,Abu24,Abu25}, while the neutron orbitals near the Fermi surface are 5/2$^-$[532], 3/2$^+$[411], 5/2$^+$[413] and 1/2$^+$[411] in 
$^{103}$Zr and $^{105}$Mo \cite{Urb09,Hua04,Din06,Sm12,Shi84,Lia95}. Therefore, two-quasiparticle configurations cannot generate angular momentum as high as $K=7$ and 9 for bands (2) and (3), given that both proton and neutron quasiparticles have $K$ values up to only 5/2. Furthermore, the band-head energies of bands (2) and (3) lie significantly below the proton and neutron pairing gaps, with 2$\Delta p\sim 1.7$ and 2$\Delta n\sim 2.1$ MeV \cite{Gue95}. Thus, these bands are unlikely to be four-quasiparticle bands. Consequently, the high $K$ values of 7 and 9 for bands (2) and (3) are attributed to collective excitations. Bands (2)–(1) and (3)–(2) in $^{104}$Nb exhibit strong linking transitions, indicating a similar origin. The enhanced $E2$ linking transitions, with $\Delta I=2$ transitions, provide further evidence for $\gamma$-vibrational quadrupole phonons.

\begin{figure}[t]
	\centering
	\includegraphics[width=\columnwidth]{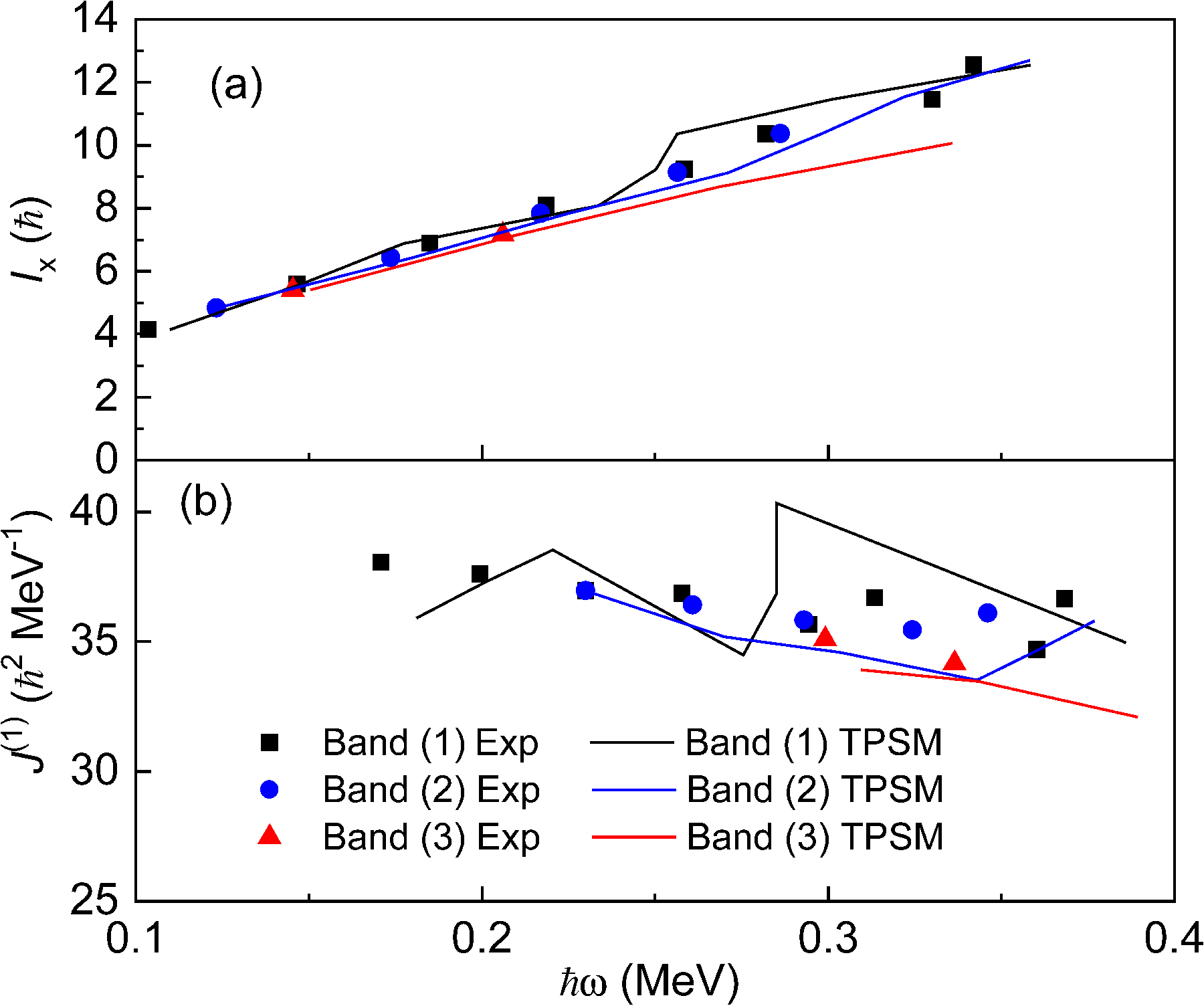}
	\caption{(a) Angular momentum alignment $I_x$ and (b) kinematic moments of inertia $J^{(1)}$ as a function of rotational frequency $\hbar\omega$ for the 0, 1 and 2 phonon $\gamma$-vibrational bands in $^{104}$Nb. The corresponding TPSM calculations are shown as continuous lines.}
	\label{Ix-J2}     
\end{figure}

The multi-phonon $\gamma$-vibrational bands in $^{104}$Nb exhibit similarities to those in $^{103,105}$Nb \cite{Wan09,Li13}.
The $\gamma$-phonon energy ratio $E_{2\gamma}$/$E_{1\gamma}$ in $^{104}$Nb is 1.81, compared to 1.79 in $^{103}$Nb and 1.97 in $^{105}$Nb. 
Using $E_I=A[I(I+1)-K^2]-E_K$ \cite{Din06},  we obtain comparable inertia parameters: $A=13.7$ for 0$\gamma$ band, 13.9 for 1$\gamma$ band and 14.5 for 2$\gamma$ band in $^{104}$Nb. 
 The angular momentum alignment $I_x$ and kinematic moments of inertia $J^{(1)}$ of bands (1), (2) and (3), shown in Fig.~\ref{Ix-J2}, exhibit a similar behavior, 
\color{black}
which further support the same origin of bands (1)-(3) as 0, 1, 2 $\gamma$ bands. As pointed out in Ref.~\cite{JS16}, the moments of inertia are the same for multi-phonon $\gamma$-vibrational bands, whereas for triaxial rotation, the moment of inertia of the $K=K_{0\gamma}+4$ band is twice that of the $K=K_{0\gamma}+2$ band. 
The average |($\mathit{g_K-g_R)/Q_0}$| for bands (1)-(3) are 0.079(3), 0.079(9), 0.083(9) b$^{-1}$, respectively, extracted from intensity ratios using the Eqs. (3) and (4) in Ref.~\cite{Wan09}. The measured angular correlation coefficients [A$_2$,A$_4$] for the 185-460 (band 1) and 185-774 (band 1-2) cascades are [0.080(27),0.012(43)] and [0.083(50),-0.013(82)], respectively. The values for 185-460 cascade yield a mixing ratio of $\delta(E2/M1)=0.34(7)$ for 185 keV indicating a $\mathit{(g_K-g_R)/Q_0}$=0.074(16). These measurements show that the present values for the 0, 1 and 2 phonon $\gamma$ bands are similar as expected.


\color{black}

\begin{figure}[t]
	\centering
	\includegraphics[width=\columnwidth]{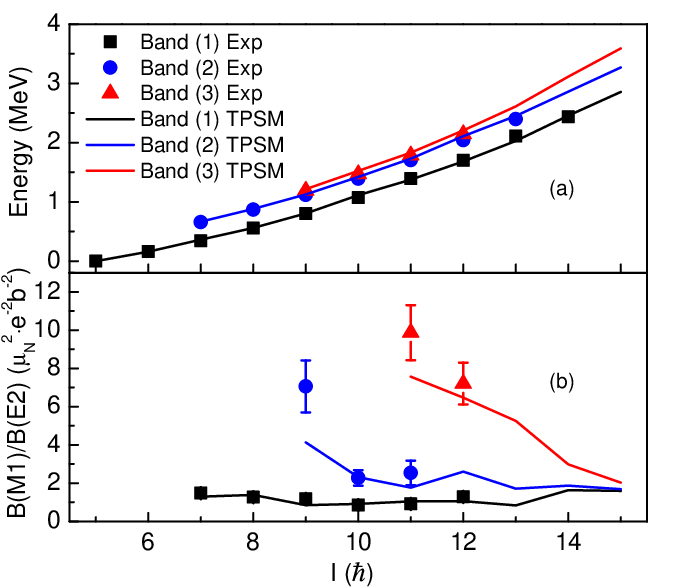}
	\caption{Energy levels and $B(M1)/B(E2)$ ratios for the 0, 1 and 2 phonon $\gamma$-vibrational bands in $^{104}$Nb as a function of spin. The corresponding TPSM calculations are shown as continuous lines (see text).}
	\label{tpsm}     
\end{figure}

%


To further investigate the $\gamma$-vibration structures of the observed bands, we have performed the triaxial projected shell
model (TPSM) analysis \cite{JS99}. This model is now widely employed to study the band structures of deformed and transitional nuclei \cite{SJ21,GH14,bh15,GH12,SJ18,JS16,Na23}. In particular, this model has provided a new insight into the nature of the observed $\gamma$-band structures and demonstrated that each triaxial intrinsic configuration projected to a given ``$K$'' value has several excited bands built on it having $K'=K+2, K+4$ and so on. These correspond to 1$\gamma$-,
2$\gamma$- and so on band structures observed in deformed and transitional nuclei.   This model needs modest computational capabilities and employs
a simplistic Hamiltonian consisting of pairing plus quadrupole-quadrupole interaction terms. 

Recently, the TPSM approach has been considerably extended for odd-odd systems \cite{SJ20} and the basis space comprises of one-neutron
coupled to one-proton, three-protons coupled to one-proton, one-neutron couple to three-protons and three-protons coupled
to three-neutrons. The deformation parameters and pairing strength used in the present work are $\epsilon$=0.32 and $\epsilon '$=0.19, 
and $G_1=20.12$ and $G_2=13.13$, respectively \cite{JA16,YS16,ZD61}.

\color{black} The lowest three calculated band structures obtained in the TPSM approach are compared with the experimental data in Fig.~\ref{tpsm} (a), and it is evident from the
figure that the calculated bands are in good agreement with the data.
\color{black}
The excited two bands with band heads at $7^-$ and $9^-$ originate from the same intrinsic state as that of the yrast-state, but
projected to $K= 7$($1\nu1\pi$)  and 9($1\nu1\pi$), respectively. 
In the traditional nomenclature, the two excited structures are categorized as 1$\gamma$- and 2$\gamma$-bands.
 In the wavefunction of $I=5$, $K=5$ $\pi$5/2$^+$[422]$\otimes\nu$5/2$^-$[532] has 91\% contribution; for $I=7$ of the $\gamma$-band, $K=7$ has a 71\% contribution; and for $I=9$ of the $\gamma$$\gamma$-band, $K=9$ has a 64\% contribution. 

The calculated in-band $B(M1)/B(E2)$ ratios using the TPSM wavefunctions are also compared with experimental values extracted from intensity ratios, as shown in Fig.~\ref{tpsm} (b).
The B(M1) values depend on the single particle states and reveal the instrinsic structure of the bands. The good agreement between the experiment and TPSM calculations
further supports the assignments of bands (2) and (3) as 1$\gamma$ and 2$\gamma$ bands based on the $5^-$ ground-state band. 
 The calculated angular momentum alignment and kinematic moments of inertia are also in agreement, as shown in Fig.~\ref{Ix-J2}.
\color{black}

In addition to the multi-phonon bands discussed above, the 141 keV isomer was also investigated to further probe the structure of $^{104}$Nb. A more precise value of the half-life of this state was measured to be 66(6) ns (inset of Fig.~\ref{spectra-all}(f)). The Internal Conversion Coefficients (ICC) of the 97 keV (band 4 to the 141 keV state) and 141 keV transitions were obtained from the measured intensity balance by gating on the transitions above and assuming E1, M1 or E2 possibilities for these two transitions. The ICC values obtained from the intensity balance for the 97 keV are 1.23(12) and 1.33(12) assuming the 141 keV as E1 and M1 multipolarity respectively. The corresponding theoretical ICC values are 0.13, 0.24, and 1.28 for E1, M1, and E2 multipolarities for the 97 keV transition \cite{bricc}. Assuming an E2 multipolarity for the 141 keV would lead to ICC of 97 keV to be much larger than that theoretical estimate for E2 multipolarity, thereby excluding this possibility. This leads to E1 or M1 multipolarity for the 141 keV isomeric transition and thus spins 4-6 for the 141 keV level. The measured 66(6) ns half-life of the 141 keV level is very large compared to the Weisskopf estimate of 8(0.1) ps corresponding to the 141-keV M1(E1) transition in $^{104}$Nb.  This half-life in $^{104}$Nb is at least an order of magnitude larger than the interband transitions in $^{101,103,105}$Nb and $^{105}$Mo. This leads to the deduced $B(M1)$ (1.2$\times$10$^{-4}$ W.u.) or $B(E1)$ (1.7$\times$10$^{-6}$ W.u.) value at least an order of magnitude lower than those in $^{101,103,105}$Nb and $^{105}$Mo. 

A possible explanation for the strongly hindered 141-keV M1(E1) transition could be the presence of a shape isomer. An oblate shape isomer at 313 keV was previously proposed in $^{109}$Nb \cite{Wat11}. QRPA calculations in Ref.~\cite{Gom21} predicted an oblate ground state for $^{104}$Nb, suggesting that configurations with an oblate shape may lie near the Fermi surface. A cranking model with the shell correction \cite{Fra00}, performed in the present work, also shows the presence of two energy minima at $\gamma$$\sim$25$\degree$ (ground) and 55$\degree$ (0.45 MeV), corresponding to triaxial and oblate shapes, respectively. Hence, this 141-keV isomer could
be an oblate shape isomer. Together with the triaxial bands (1)–(3), $^{104}$Nb may present a case of shape coexistence. A more hindered transition between triaxial-oblate shape coexistence is expected in $^{104}$Nb. This possible shape coexistence in this odd-odd nucleus could be attributed to the influence of valence protons and neutrons driving the nucleus toward different shapes in this A region.

In conclusion, the first multi-phonon $\gamma$-vibrational bands in an odd-odd nucleus are reported in $^{104}$Nb. This challenging measurement was only possible due to the combination of two complementary $\gamma$-ray spectroscopic measurements. Two new bands with similar structures are assigned as 1 and 2 phonon $\gamma$-vibrational bands coupled to the yrast band, with their configurations strongly supported by the good agreement of the experimental data with TPSM calculations. The known isomer at 141 keV is proposed to be an oblate shape isomer, and along with triaxial shape of the $\gamma$-band indicates a possible shape coexistence in $^{104}$Nb. Investigations into more neutron-rich Nb isotopes could throw more light on the role of the valence neutron and proton in the manifestation of multi-phonon bands. The search for the presence
of multi-phonon bands in odd-odd nuclei $A\sim 130$ and $\sim$230,
could provide further insight into the role of the unpaired nucleon in different sets of orbitals on $\gamma$-vibrations in nuclei. This could help to better understand the driving force of valence neutrons and protons and the dynamical behaviour of surface vibrations, and also the nature of vibration modes in other regions of physics.

\section*{Acknowledgements}

We would like to thank the GANIL staff for their technical contributions. We would also like to thank the e661 and e680 collaborations and gratefully acknowledge the AGATA collaboration for the availability of the AGATA $\gamma$ tracking array at GANIL.  The authors would like to thank N. Gavrielov and A. Aprahamian for their useful comments and  carefully reading  the manuscript, and Q.B. Chen for helpful theoretical discussions. The work at Shandong University is supported by the National Natural Science Foundation of China No. 12225504 and No. 12475123. The work at Vanderbilt University and Lawrence Berkeley National Laboratory is supported by the U.S. Department of Energy under Grant No. DE-FG05-88ER40407 and Contract No. DE-AC03-76SF00098. The AGATA collaboration acknowledges support from the European Union FP7th Integrating Activity ENSAR, Contract No. 262010. SB acknowledges the support received from CEFIPRA project No. 5604-4 and LIA France-India agreement. YHK acknowledge the support from the Institute for Basic Science (IBS) of the Republic of Korea (Grants No.~IBS-R031-D1). EW acknowledges support from Shandong Provincial Outstanding Young Scholars Fund (Overseas) Program No. 2025HWYQ-028 and Shandong University Qilu Young Scholar.

\end{document}